\documentstyle[12pt]{article}

\begin{document}

\begin{center}
{\Large {\bf Tidal Tails and Galaxy Evolution}}
\end{center}

\begin{center}
{\large V.P.~Reshetnikov, N.Ya.~Sotnikova} \\
{\it Astronomical Institute of St.Petersburg State
University, \\ 
Bibliothechnaya pl. 2, Petrodvoretz, 198504 S.-Petersburg, Russia}
\end{center}

\abstract{We review recent results on the tidal structures of spiral
galaxies. Topics included are general characteristics of tails; kinematics
of tidal structures and dark haloes of host galaxies; frequency of tidal
distortions at $z \sim 1$.} 

\vspace{0.8cm}

{\bf Introduction}

Tidal features are very old known extragalactic objects. It is not generally
realized that first remark on them has more than 2 hundred years. William
Herschel was the first who described several double and multiple systems of
faint nebulas and noted that some of the nebulas are connected by thin
strips of luminous matter (see [1]). Almost two hundred years later, Toomre
\& Toomre [2] demonstrated by means of numerical modeling that such exotic
and strange objects can be naturally explained as a tidal distortions of
gravitationally interacting galaxies.

Extended tidal structures are almost unexplored objects from both
observational and theoretical points of view. We have now detailed
information about 10 or 20 such structures only. In our report we shall
discuss shortly general observational characteristics of the tails and
stress the importantance of their investigation.\\

{\bf General characteristics of tidal tails}

{\bf 1) Local frequency}

The local fraction of tailed objects is about (1-2)\% of all galaxies [3,4].
This estimation is obtained for relatively bright optical tails. While decreasing
surface brightness level we shall observe even larger fraction of tidal
distortions. So tidal tails are rare but not extremely infrequent features.

{\bf 2) Spatial extent}

The typical sizes of tails are, on average, comparable with the sizes of
host galaxies. But in some cases we observe huge structures -- with lengths
reaching hundreds kiloparsecs (for instance, a 180 kpc tail in Arp~299
[5]!).

{\bf 3) Surface brightness}

Known tidal features are generally very faint -- with surface brightness
levels in the $B$ passband around 24$^m$--25$^m$ [6,7].

{\bf 4) Total luminosity}

Surface brightnesses of the tails are low but they are very extended and,
therefore, total luminosity may be significant, with mean value about
quarter of the luminosity of main galaxy [6].

{\bf 5) Optical colors}

Optical colors are, on average, bluer than those for the main galaxies. On
the whole, the mean color indices are close to those for late-type spiral
galaxies: $B-V \approx +0.5$ [6,7].

{\bf 6) m(HI), m(H$_2$)/m(HI)}

Tidal tails are usually gas-rich structures with typical HI masses exceeding
or equal 10$^9$ solar masses [8]. In several cases the tails contain even
most part of HI gas associated with whole galaxy. Molecular gas (CO emission)
was discovered recently in the tails of two interacting galaxis [9]. The mass
ratio of the molecular gas to HI for those two tails is typical for
late spirals. \\

{\bf Simple physics}

To understand the development of tidal tails, one must recall how the
water surface of the oceans get stretched radially by differential
gravitational attraction exerted on
it by our Moon. The differential forces between near and far side of the
Earth depend on the third power of the Moon's distance from Earth. That is
why the oceans tides are rather mild. When two galaxies experience a close
encounter with perigalactic distance is compared with galaxies' sizes, the
tidal field of one galaxy stretches its neighbour radially and then the
galaxies' rotation shears off stars and gaseous clouds from outskirts of
their parent galaxies. As a result, stars on the far side of each disk is
ejected into long and thin tails.

The change of $i-$th star momentum ($\Delta v_i$) during the encounter
($\Delta t$) can be expressed in term of perigalactic distance and relative
velocity $v_{\rm rel}$.
$$
\Delta v_i = \frac{F_{\rm tid}}{m_i} \Delta t \propto
\frac{1}{r_{\rm per}^3} \frac{r_{\rm per}}{v_{\rm rel}}
\propto \frac{1}{r_{\rm per}^2 \, v_{\rm rel}}  \, .
$$
The additional energy may be sufficient to transfer the star from position
$r$ to position $r + \Delta r$
$$
\frac{(\Delta v_i)^2}{2} = \varphi(r_i + \Delta r) - \varphi(r_i) \approx
\left| \frac{d \varphi}{d r} \right| \Delta r \, ,
$$
where $\varphi$ - is gravitational potential. Then the length of the tail
can be estimated as
\begin{equation}
l_{\rm tail} \approx \Delta r \propto
\frac{1}{\displaystyle r_{\rm per}^4\,v_{\rm rel}^2\,\left|\frac{d\varphi}{dr}\right|}
\, .
\label{long}
\end{equation}
One can see from Eq.\ref{long} that the extent of tidal tails depends
strongly on resonances between the rotational and orbital motions
(prograde encounters, in which the directions of galaxies rotation and
orbital motion are the same -- $v_{\rm rel}$ is small, -- are most effective
for tail building), on perigalactic distance and on the deepness of the
common gravitational well.

And now when we know some general things about tails, let's discuss
why it is so interesting and important to study them. We shall discuss
only three major points related to essential questions of
galaxy evolution. \\

{\bf Star formation in unusual environment}

Tidal tails are very unusual settings for star formation. There are several
groups of arguments in favour of ongoing star formation in some of such
features. For instance, they are gas-rich and have blue optical colors
typical for late spirals and spiral arms. Bright
blue star forming knots, often associated with HI condensations, are
observed in several extended tails. For instance, we found a number of
H$\alpha$ condensations in the
straight and long ($\sim 40$ kpc) tail of NGC~4676 (The Mice) [10].
Characteristics of the condensations (linear sizes, optical and H$\alpha$
luminosities) are common for giant HII complexes.
The star formation rate in the tail
estimated on the total H$\alpha$ luminosity
is typical for the disks of normal spirals
(10$^{-9}$M$_{\odot}$/yr/pc$^2$). Analytical estimations and numerical
calculations indicate that the main star formation mechanism in the tidal
tail of the Mice is large-scale gravitational instability in the gas of the
tail.

Last years observational similarity of the brightest parts of tails and dwarf spiral
galaxies have restored an old Zwicky's idea that dwarf
galaxies can be formed from the tidal material ejected during the
interactions.
And now this idea became an interesting and actively growing field of study.
For instance, as the interaction and merger rate was higher in the past, the
production of such Tidal Galaxies might be very large. This mechanism can
explain in part an excess of the faint blue galaxies. \\

{\bf Tidal tails and dark haloes}

Another interesting field of work is the constraints on galactic
dark halos characteristics following from the kinematics and morphology
of tails. The extent of the tidal tail is very sensitive to the
global dynamical structure of the interacting galaxies [11,12] --
it is difficult for long, massive tails to form in galaxy
collisions in which the progenitors are surrounded by very
large halos. As one can see from Eq.\ref{long} the extent of the tail is
inversly proportional to the deepness of gravitational well, and tails
extracted from disks might be unable to climb out of deep halo potential
wells. But we observe very extended tails! It is suggested that this fact
restricts the dark to luminous matter ratio in galaxies [11].

Such statistical approach have several evident limitations.
In our work [10] we proposed to constrain dark haloes
through detailed modelling of specific interacting systems
with good observational data about central galaxies and tails.
Especially we need good kinematical data for the tails but,
unfortunately, such data are extremely poor now.

Probably, the best studied case is well-known system The Mice.
With the 6-m telescope we traced emission-lines rotation curve of
the northern edge-on tail up to 40 kpc from the nucleus [10].
Using all available
information about the central galaxies, we have performed
the numerical modelling of the system.
We found that the observed high radial velocities in the tail can be
explained {\it only} if the system members possess rather massive dark
halos -- with dark to luminous mass ratio within the region up to the tip of
the tail about 4.

Besides the Mice, we have the results of kinematical observations
for several other objects with tails (work in progress). \\

{\bf Interaction rate evolution}

Tidal deformations are clear indicators of recent gravitational
perturbations of galaxies. Therefore, one can use their statistics
to estimate the interaction rate in the past. It is very
important because the rate of interactions and mergers
at different redshifts depends on the model of the Universe,
and on the details of galaxy formation models.

Tidal tails are faint structures. Due to cosmological dimming
and $K$-correction, we can observe such distortions out to
moderate redshifts only (1 or 1.5). Using the deepest currently
available fields -- North and South Hubble Deep Fields, -- we
selected 25 galaxies with probable tails at redshift between 0.5
and 1.5 [4]. General characteristics of the suspected
tails are, on average, the same as in local interacting galaxies.
We found that volume density of galaxies with tidal tails changes
with z as $(1+z)^4$. Therefore, we estimated
the rate of close encounters between galaxies of comparable
mass leading to the formation of extended tidal structures.
If this rate reflects the merger rate, our data support a
steeply increasing merger rate at z$\sim$1 and are consistent
with current theoretical expectations.

Finally, we would like to stress that further detailed study and modelling
of beautiful tidal features will be a powerful tool to investigate
many important question of extragalactic astronomy, such as
star formation process, the mass and extent of galactic haloes,
interactions at high redshifts and so on. \\

{\bf References}\\
1. Eremeeva, A.I. {\it Herschel's Universe}, Nauka, Moscow, 1966 \\
2. Toomre, A. and Toomre, J., 1972, {\it Ap.J.} {\bf 178}, 623 \\
3. Karachentsev, I.D. {\it Binary Galaxies}, Nauka, Moscow, 1987 \\
4. Reshetnikov, V.P., 2000, {\it A\&A} {\bf 353}, 92 \\
5. Hibbard, J.E. and Yun, M.S., 1999, {\it A.J.} {\bf 118}, 162 \\
6. Schombert, J.M., Wallin, J.F., Struck-Marcell, C., 1990,
{\it A.J.} {\bf 99}, 497 \\
7. Reshetnikov, V.P., 1998, {\it Pis'ma v AZh} {\bf 24}, 189 \\
8. Hibbard, J.E. and van Gorkom, J.H., 1996, {\it A.J.} {\bf 111}, 655 \\
9. Braine, J., Lisenfeld, U., Duc, P.-A., 2000, {\it Nature} {\bf 403},
867 \\
10. Sotnikova, N.Ya. and Reshetnikov, V.P., 1998, {\it Pis'ma v AZh}
{\bf 24}, 97 \\
11. Dubinski, J., Mihos, J.Ch., Hernquist, L., 1996, {\it Ap.J.}
{\bf 462}, 576 \\
12. Dubinski, J., Mihos, J.Ch., Hernquist, L., 1999, {\it Ap.J.}
{\bf 526}, 607 \\

\end{document}